\documentclass[useAMS,usenatbib,usegraphicx]{mn2e}
\usepackage{epsfig}
\newcommand{\etal}{{\it et~al.\/\ }}
\newcommand{\kms}{km~s$^{-1}$}
\newcommand{\Msun}{M$_{\sun}$}
\title[Galaxy Cores in $\Lambda$CDM]{Cuspy No More: How Outflows Affect the Central Dark Matter and Baryon Distribution in  $\Lambda$CDM Galaxies.}
\author[F. Governato et al.]{F.Governato$^{1}$\thanks{E-mail:(FG); fabio@astro.washington.edu} 
        A.Zolotov$^{2}$,
        A.Pontzen$^{3}$,
        C.Christensen$^{4}$,
        S.H.Oh$^{5,6}$, 
        A.\,M.Brooks$^{7}$,\newauthor
        T.Quinn$^{1}$,
        S.Shen$^{8}$,
        J.Wadsley$^{9}$
\\
$^{1}$Astronomy Department, University of Washington, Box 351580, Seattle, WA, 98195-1580 \\
$^{2}$Racah Institute of Physics, Hebrew University , Jerusalem, 91904, Israel \\
$^{3}$James Martin Research Fellow, Oxford Astrophysics, Univ. of Oxford, Denys Wilkinson Bldg., Keble Road, OX1 3RH, UK\\
$^{4}$Theory Fellow, Astronomy Department, Univ. of Arizona, Tucson\\
$^{5}$International Centre for Radio Astronomy Research (ICRAR), Univ. of Western Australia, 35 Stirling Highway, Perth, WA 6009, AU\\
$^{6}$ARC Centre of Excellence for All-sky Astrophysics (CAASTRO)\\
$^{7}$Grainger Postdoctoral Fellow, Department of Astronomy, University of Wisconsin\\
$^{8}$Department of Astronomy \& Astrophysics, UC Santa Cruz\\
$^{9}$Dept. of Physics and Astronomy, McMaster Univ., Hamilton, Ontario, L88 4M1, Canada
}

\begin{document}

\date{Submitted 2012 January 8th.}

\pagerange{\pageref{firstpage}--\pageref{lastpage}} \pubyear{2002}

\maketitle

\label{firstpage}

\begin{abstract}

  We examine the evolution of the inner dark matter (DM) and baryonic
  density profile of a new sample of simulated field galaxies using
  fully cosmological, $\Lambda$CDM, high resolution SPH + N-Body
  simulations. These simulations include explicit H$_2$ and metal
  cooling, star formation (SF) and supernovae (SNe) driven gas
  outflows.  Starting at high redshift, rapid, repeated gas outflows
  following bursty SF transfer energy to the DM component and
  significantly flatten the originally `cuspy' central DM mass profile
  of galaxies with present day stellar masses in the 10$^{4.5}$--
  10$^{9.8}$ $M_{\odot}$ range.  At z$=$0, the central slope of the DM
  density profile of our galaxies (measured between 0.3 and 0.7 kpc
  from their centre) is well fitted by $\rho_{DM}$ $\propto$
  r$^{\alpha}$ with $\alpha \simeq -0.5 + 0.35 \log_{10} \left(
    M_{\star}/ 10^8\,M_{\odot} \right)$ where M$_{\star}$ is the
  stellar mass of the galaxy and 4 $<~\log M_{star}~< 9.4$.  These
  values imply DM profiles flatter than those obtained in DM--only
  simulations and in close agreement with those inferred in galaxies
  from the THINGS and LITTLE THINGS survey.  Only in very small halos,
  where by z~$=$~0 star formation has converted less than $\sim$
  0.03\% of the original baryon abundance into stars, outflows do not
  flatten the original cuspy DM profile out to radii resolved by our
  simulations. The mass (DM {\it and} baryonic) measured within
    the inner 500 pc of each simulated galaxy remains nearly constant
    over four orders of magnitudes in stellar mass for M$_{star}$
    $<$~10$^{9}$ M$_{\odot}$. This finding is consistent with
    estimates for faint Local Group dwarfs and field galaxies.

    These results address one of the outstanding problems faced by the
    CDM model, namely the strong discrepancy between the original
    predictions of cuspy DM profiles and the shallower central DM
    distribution observed in galaxies.
 \end{abstract}

\begin{keywords}
Galaxies: formation -- Cosmology -- Hydrodynamics.
\end{keywords}

\section{Introduction}

The predictions of the $\Lambda$CDM cosmological model are in
excellent agreement with observations of the assembly of cosmic
structures on large scales \citep{eke96, riess98, spergel07}. The
observed properties of dwarf galaxies, however, have presented strong
challenges to the model at galactic scales. DM-only simulations
predict that DM halos should follow a quasi--universal Einasto profile
\citep{navarro96,moore98b,reed05a,maccio09,stadel09,navarro10}, defined
by a power law density profile $\rho$ $\propto$ r$^{\alpha}$ with
$-1.5<\alpha <-1$ in their central regions. Such simulations therefore
predict steep (or ``cuspy'') inner density profiles. Observations of
small galaxies ($V_{peak} \sim 30-60$ km/s), however, have repeatedly
shown that their central DM profiles are shallower than the
predictions of DM only simulations at scales of 0.1-1 kpc
\citep[e.g.,][]{flores1994,moore98b,swaters03,blitz05,deblok08,donato09,
primack09, kuzio10, oh10a, elson10}.  In these observational works the measured
$\alpha$ slopes range between 0 and -1.  (Note: in this paper we will
identify halos with DM profiles shallower than NFW or Einasto as
'cored').  These observational results provided a long standing
challenge to CDM at scales that cannot probed by CMB or Ly$\alpha$
experiments \citep{croft02}. They also prompted the development of
several alternative DM models such as Warm DM (WDM), meta--DM and Self
Interacting DM (SIDM)
\citep{spergel00,dave01,knebe02,ahn05,strigari07,colin08,martinez09,abdo10,
loeb11,lovell11,koda11,vogelsberger12} and alternative gravity theories such as MOND
\citep{mcgaugh05,gentile11}.

There is emerging evidence that our poor understanding of the baryonic
processes involved in galaxy formation is the source of the
inconsistency between the observations of dwarf galaxies and the
predictions of CDM.  Models of the effect of feedback on the
  structure of galaxies and the efficiency of SF were originally
  motivated by the discrepancy between the observed number of
  dwarf galaxies and the much higher number density of small DM halos
  predicted by CDM models \citep{dekelsilk86,moore98,bower06}.
Without requiring a major change to the $\Lambda$CDM scenario, several
models have been presented in the past advocating the evolution of
originally cuspy CDM halos into halos with 'cored' density profiles
through SN feedback or dynamical friction
\citep{navarro96cores,dekelcores03,dekelcores2,mo04,
  read05,mashchenko06,goerdt06,tonini2006,elzant2004,
  romanodiaz09,delpopolo09,lackner10,goerdt10,desouza11,pontzen11}.  In
\citet[][hereafter, G10]{G10} we presented self-consistent, DM + gas
dynamic simulations where shallow DM cores arise naturally in a CDM
cosmology \citep[see also][]{maccio11}. In these simulations, energy
feedback from SNe in star forming regions generate repeated, fast gas
outflows. These outflows efficiently remove gas from the inner kpc of
protogalaxies \citep{brook11} and, in smaller galaxies, are able to
expel a large fraction of gas from the galaxy altogether.

In \cite{pontzen11} we presented a coherent analytical
model for core formation that correctly matches results from our
simulations.  In this model multiple, rapid gas outflows transfer
energy to the collisionless DM and create DM cores of about 1 kpc in
size in halos of total mass $2-3 \times 10^{10} M_{\odot}$. As also
described in detail in G10, a crucial ingredient is the spatial
distribution of the SF events.  If SF is allowed in low density gas
$\rho \sim 0.1-1$ amu/cm$^3$ (typical in most previous, lower resolution
simulations) outflows are weaker and do not generate cores, even if
the feedback scheme remains the same. The necessity  of simulations to
resolve dense gas regions calls for simulations of very high mass and
spatial resolution \citep{mashchenko08,saitoh08,ceverino09}.

As a key step toward understanding the properties of DM halos at very
small masses, many faint Milky Way dwarf satellites have recently been
discovered, some with less than one millionth of the Milky Way's
luminosity \citep[e.g.,][]{willman05, belokurov07}. The first
kinematic studies of `ultra-faint' dwarf galaxies have measured mass
to light (M/L) ratios as low as 1000 \citep{simon07,geha09}. These
results suggest that the smallest cosmic structures where star
formation took place might have been identified. It has been argued
that the observed population of faint galaxy satellites should be
hosted inside cuspy halos, as small, cored halos would easily be
destroyed by the Milky Way tidal field \citep{penarrubia10}. However,
estimates of the mass profiles of ultra--faint dwarfs are uncertain
and critically depend on the detailed kinematic of their stars, which
are used as dynamical tracers \citep{walkerjorge11}. Unlike most field
dwarfs, where the evidence for DM cores is robust, current
observations have not securely determined if low mass satellites of
the MW are hosted inside cored or cuspy DM halos, although
\cite{walkerjorge11} and \cite{jardel11} reported that the Sculptor
and Fornax dwarf galaxies might indeed have cores.

Surprisingly, while the Milky Way low luminosity dwarfs span five
orders of magnitude in luminosity, dynamical studies estimate that
these galaxies contain the same total (baryons$+$DM) mass in their
innermost part, about 10$^7$ M$_{\odot}$ within the inner 300pc
\citep{strigari08}. This result provides a useful constraint for all
models of galaxy formation and can be interpreted assuming a
truncation in the halo mass function as predicted in WDM models
\citep{avila01,bode01,boylan11,parry11,schneider11} or to the
suppression of SF in halos with virial temperature less than the
cosmic UV background \citep{quinn96,benson02,stringer10b}.  Both processes
would set a minimum halo scale for star formation, hence a common mass
at a fixed radius. Feedback and gas heating from cosmic sources have
been shown \cite[among many]{g07,font11}, to alleviate the above
mentioned overabundance of sateliites in CDM, while WDM faces already
significant observational constraints \citep{viel05,kuzio10,lovell11}
and several works presented evidence that WDM halos would form cores
much smaller than observed, \citep{dalcanton01,knebe02,strigari07},
making this model less attractive.

A few models have tried to explain the observed flat stellar mass -
central mass relation measured in Strigari et al (2008), while
assuming cuspy DM halo profiles \citep{rashkov11}. Some
\citep{li09,maccio09} invoke a large scatter in star formation
histories (SFH) and halo assembly histories of dwarf galaxies to
explain the observed flat luminosity-central mass relation over a
large range of halo masses. These results however, are based on
DM-only simulations paired to semi analytical models.  They therefore
neglect important interactions between the baryonic and the DM
component of halos, as well as the effect of H$_2$ cooling on SF in
dense regions \citep{gnedin09}.  Ongoing surveys such as ANGST
\citep{ANGST09}, THINGS \citep{walter08} and LITTLE THINGS
\cite[Hunter et al 07]{zhang11} are providing data on the central mass
distribution of {\it{field}} galaxies with V$_{peak}$ $<$ 60 km/sec
and provide new, strong constraints on the central mass distribution
of galaxies \citep[][hereafter OH11]{ohsim11} that need to be properly
taken into account. In particular, LITTLE THINGS and THINGS are able
to measure DM profiles over a range of galaxy stellar masses and to
evaluate the total mass within the central region of galaxies more
massive than those in the original 'Strigari relation'.

To summarize, observational measurements of the central mass
distribution of galaxies and  of the central slope of
the underlying DM profile as a function of a galaxy stellar mass, may
 shed light on the nature of DM and of baryon/DM
interactions at scales much smaller than those probed by cosmological
test.

In this work we focus on {\it field} galaxies and their structural
properties, and compare the highest resolution set of simulations of
small galaxies to date with measurements of DM slopes obtained from
new data from the THINGS and LITTLE THINGS surveys. Focusing on field
galaxies allows us to separate the effects of gas outflows from those
of tidal interactions \citep{mayer01}. The goal of the analysis
presented here is to evaluate if galaxies formed in a CDM cosmology
have central DM and baryon distributions consistent with observations
once the effect of realistic gas outflows is evaluated in a full
cosmological context.  To achieve this goal we study a new set of high
resolution galaxies formed in a full $\Lambda$CDM cosmological
context.  The simulations presented here include a consistent
implementation of metal line cooling and H$_2$ physics (Christensen et
al '11, submitted, CH11 hereafter), essential to correctly model SF in
low metallicity gas and in low mass halos \citep{li09,krumholz11}. The
highest resolution simulations in our sample resolve individual SF
regions as small as a few $10^{4} M_{\odot}$.  The simulations have
not been tuned to produce cores, but rather to form a realistic amount
of stars and to reproduce the stellar mass/halo mass relation
\citep[Munshi et al, in prep]{conroy06,tollerud11}.  As a test of our
predictions for the central mass distribution of galaxies, we also
show that simulations naturally reproduce an updated flat stellar mass
- central mass relation obtained combining MW and field dwarf
galaxies.  We describe the simulations and the THINGS data in \S2, the
results on the shape of the DM profiles in \S3 and the results on the
central mass - stellar mass relation in \S4. In \S5 we conclude.

\begin{table*}
 \centering
 \begin{minipage}{140mm}
  \begin{tabular}{@{}lccccccc@{}}
  \hline
Simulation &  Galaxies stellar  &  DM part.  & 
 Star part.  &  Softening   &  Overdensity & 
 Particles &  V$_{peak}$ \\
 & masses \Msun & mass \Msun & mass (\Msun)  & (pc) & $\Delta \rho /\rho$ &  within R$_{vir}$ & \kms \\
 \hline
Fields 1 \& 2       &  10$^{10}$-10$^8$         & 1.6$\times$10$^{5}$  & 8 $\times $10$^{3}$ & 170 & 0.38 -- 0.03  & 3.4-0.05$\times$10$^{6}$  & 100-40 \\
Field 3 \& 4           &  3 $\times$ 10$^8$-10$^5$ & 2$\times$10$^{4}$   & 10$^{3}$             & 85 & 0.58 -- -0.07   & 2-0.05$\times$10$^{6}$    &  55-30 \\
Field 5             &  10$^8$-10$^{3.5}$            & 6$\times$10$^{3}$   & 4.2$\times$10$^{2}$  & 64 & 0.01   & 2-0.05$\times$10$^{6}$    &  35-10 \\
\hline
\end{tabular}
\end{minipage}
\caption{{\it Properties of the simulated galaxies.} All masses in M$_{\odot}$. 
Column (2)  lists the  total stellar mass range for each galaxy in the subsample  at $z$=0.  
Columns (3) and (4) list the mass  of individual dark matter and star particles, 
respectively.  Column (5) shows  $\epsilon$, the spline gravitational force softening, in pc. Column (6) shows the overdensity in units of the average density around the most massive halo in that zoomed-in region  measured on a scale of 4h$^-1$ Mpc.  (7) lists the range in total number of particles (gas, stars and DM) within 
the virial radius of the halo at $z$=0. (8) gives the peak velocity at z$=$0. All simulations but Field 4 have also been run as DM-only (See Fig.2).}
\end{table*}

\section{Simulations and Observational Data}
\label{sims}

The simulations used in this work were run with the N-Body + Smoothed
Particle Hydrodynamics (SPH) code {\sc Gasoline}
\citep{wadsley04,stinson06} in a fully cosmological $\Lambda$CDM
context:

$\Omega_0=0.26$, $\Lambda$=0.74, $h=0.73$, $\sigma_8$=0.77, n=0.96.
The galaxies were originally selected from two uniform DM-only
simulations of 25 and 50 Mpc per side.  From these volumes five
field--like regions where selected, each centered on a galaxy sized
halo of different mass (3 $\times$ 10$^{11}$, 2 $\times$ 10$^{11}$, 3
$\times$ 10$^{10}$, 2 $\times$ 10$^{10}$ and 10$^{10}$ M$_{\odot}$).
Each region was then resimulated at higher resolution and with baryons
using the `zoomed-in' volume renormalization technique
\citep{katz93,brooks07,pontzen08}, while fully preserving the surrounding
large scale structure. This technique allows for significantly higher
resolution while capturing the effect of large scale torques that
deliver angular momentum to the galaxy \citep{barnes87}.  The mass
overdensity $\delta \rho$/$\rho$ for each chosen field ranges from -0.07 to
0.58 when measured on a scale of 4h$^{-1}$ Mpc (see table 1).

To simulate even very small halos with million of resolution elements,
the mass and spatial resolution of each zoomed region are inversely
proportional to the mass of the largest halo in each one of them (see
Table 1).  The force spline softening ranges between 64 and 170 pc in
all runs and it is kept fixed at z $<$ 10.  Star particles are formed
with a mass of 400-8000 M$_{\odot}$. The gas smoothing length is
allowed to shrink to 0.1 times the gravitational softening in very
dense regions (resulting values around 0.5 are typical) to ensure that
hydro forces dominate at very small scales. The main galaxy in every
zoomed region contains several million particles within its virial
radius (R$_{vir}$, defined as the radius at which the average halo
density = 100 $\times$ $\rho_{crit}$). Every zoomed region also
contains several smaller galaxies with identical mass and spatial
resolution, which we include in our analysis.  With this approach the
total high resolution sample contains 15 field galaxies with between
several million to 50000 DM particles within R$_{vir}$ covering a halo
mass range of 3 orders of magnitude. Galaxies and their parent halos
were identified using AHF\footnote{{\bf{A}}MIGA's {\bf{H}}alo
  {\bf{F}}inder, available for download at
  http://popia.ft.uam.es/AMIGA/} \citep{knollmann09}. Similar to
criteria used in previous works one galaxy undergoing strong
interactions at z$=$0 was excluded from the sample.  In this work we
follow up on G10 and OH11, where we studied the stellar and DM content
of halos M $\sim$ 10$^{10}$ M$_{\odot}$ and extend our predictions on
a larger range in halo masses (from a few times 10$^8$ to 3 $\times$
10$^{11}$ M$_{\odot}$), peak velocities V$_{peak}$ (10 to 100 km/sec)
and stellar masses M$_{star}$ (from 10$^4$ to almost 10$^{10}$ M$_{\odot}$),
spanning a range of halo spin values and accretion histories. In
observational terms this range covers faint dwarfs to normal field
disk galaxies \citep{geha06}.  No halos in our sample have been
contaminated by particles from the lower resolution volumes and the
zoomed in regions are large enough (a few Mpc in each case) to contain
all the gas expelled from galaxies by outflows (that can reach as far
as several times the virial radius of the parent galaxy).  This
approach allows us to test for the effects of resolution over a large
range in halo masses (i.e. Fig.1 and Fig.2 contain halos of
overlapping mass but ran at different resolutions) Given the high
resolution of our simulations star forming regions as small as 10$^5$
M$_{\odot}$ are identified by hundreds of gas and star particles.

As a significant improvement over most cosmological simulations
carried to the present time these new simulations include both metal
lines and H$_2$ cooling \citep[CH11]{shen10}. The simulations include
a dust dependent description of H$_2$ creation and destruction by
Lyman Werner radiation \cite[CH11]{gnedin09}. Metal lines cooling, the
H$_2$ fraction, and self shielding of high density gas from local
radiation play an important role in determining the structure of the
ISM and where SF can occur \citep{kennicutt98,
  elmegreen08,krumholz05,bigiel08, gnedin09, gnedin11,narayanan11}. With this
approach the {\it local} SF efficiency is linked directly to the local
H$_2$ abundance, as regulated by the gas metallicity and local
radiation from young stars. As a result, in our simulations stars
naturally form in high density regions around 10-100 amu cm$^{-3}$
without having to resort to simplified approaches based on a fixed
local gas density threshold \citep{G10,saitoh08,eris11,kuhlen11}. A
Kroupa (1993) IMF and relative yields are assumed.  We include a gas
heating spatially uniform, time evolving UV cosmic background
following an updated model of \citet{haardtmadau96}. Gas heating from
UV radiation progressively suppresses star formation in galaxies below
10$^{10}$ M$_{\odot}$, making small DM halos completely void of stars
\citep{benson02} and reducing the overabundance of dwarf satellite
galaxies \citep{moore98}.  The smallest galaxies in our sample have some SF
occurring before reionization (z $\sim$ 9 in our model) likely
associated to H$_2$ cooling and then in small, sparse bursts
thereafter.

The full details of our physically motivated SN feedback
implementation and its applications have been described in several
papers and shown to reproduce many galaxy properties over a range of
redshifts:
\citet{stinson06,brooks07,g07,pontzen08,g09,zolotov09,pontzen10,brook11,brooks11,eris11}.
As in G10 The SFR in our simulations is set by the local gas density
$(\rho_{gas})^{1.5}$ and a SF efficiency parameter, $c_* =$ 0.1 to
give the correct normalization of the Kennicutt-Schmidt relation (the
SF efficiency for each star forming region is much lower than the
implied 10\%, as only a few star particles are formed before gas is
disrupted by SN winds). The maximum temperature for gas to turn into
stars is set to 3000K and the efficiency of SF is then further
multiplied by the H$_2$ fraction, which effectively drops to zero in
warm gas with T $>$ 10,000 K. As massive stars evolve into SN, mass,
thermal energy and metals are deposited into nearby gas particles. Gas
cooling is turned off until the end of the snow plow phase as
described by the Sedov-Taylor solution, typically a few million years.
The amount of energy deposited amongst those neighbors is 10$^{51}$
ergs per SN event.  Energy deposition from SN feedback leads to
enhanced gas outflows that remove low angular momentum gas from the
central regions of galaxies \citep{brook11}.  We have verified that in
this set of simulations the `loading factor' of the winds, i.e. the
amount of baryons removed is typically a few times the current SFR,
similar to what is observed in real galaxies over a range of redshifts \citep{martin99,shapley03,kirby11,vanderwel11}.

\begin{figure}
\includegraphics[angle=0,width=88mm]{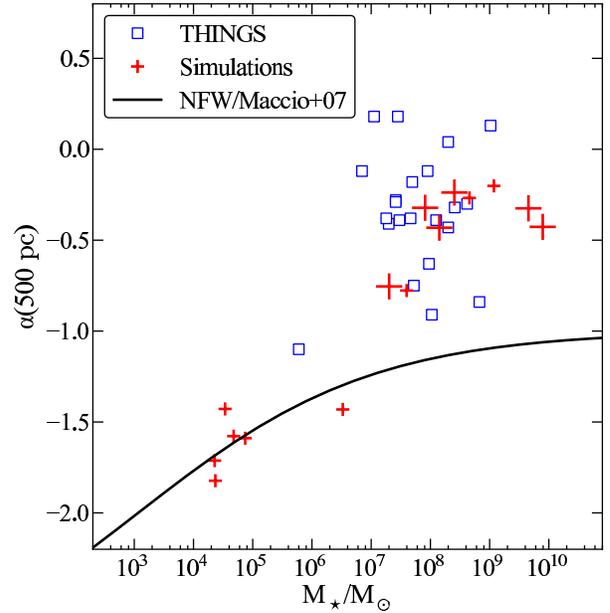}
\caption{ {\it The slope of the dark matter density profile}
 $\alpha$ vs stellar mass measured at 500 pc and z=0 for all the
 resolved halos in our sample. The Solid 'DM-only' line is the slope
 predicted for the same CDM cosmological model assuming i) the NFW 
 concentration parameter trend given by Macci\'o et al (2007)
 and ii)  the same stellar mass
 vs halo mass relation as measured in our simulations to convert from
  halo masses. Large Crosses: haloes resolved with more than 0.5
  $\times$ 10$^6$DM particles within R$_{vir}$. Small crosses: more
  than 5 $\times$ 10$^4$ DM particles. The small squares represent 22
  observational data points measured from galaxies from the THINGS and
  LITTLE THINGS surveys.
}
\label{fig1}
\end{figure}

As SF is limited by the local H$_2$ abundance, stars form only in high
density peaks sufficiently shielded from radiation from hot stars and
The SFHs of the galaxies in our simulated sample are bursty over a
significant fraction of the Hubble time, but especially at high
redshift whey each galaxy is still divided into individual
progenitors. Bursty phases typically last 10-100 Myrs with SFR
variations on shorter timescales and SF enhanced by a factor of $\sim$
4-20, similar to what measured in Local Group dwarfs
\citep{mcquinn10}.   As discussed in Pontzen \& Governato
(2011) a bursty SF is necessary to create the fast outflows able to
transfer energy to the DM component.  Outflows also decrease the SF efficiency
 in halos with total mass smaller than a few 10$^{10}$ \Msun.
In our set of simulations outflows are predominant at
high-z\footnote{http://youtu.be/FbcgEovabDI?hd=1} when SF peaks and
galaxy interactions are common. These outflows affect the halos that
will subsequently merge to form the central regions of the final,
present day galaxies.

In the mass range explored by our simulation (up to halos with
M$_{vir}$ = 3 $\times$ 10$^{11}$) the ratio of stellar mass/halo mass
(the SF efficiency) is a strong function of halo mass, roughly scaling
as M$_{stars}$ $\propto$ M${_{vir}}^2$. In this halo range SF becomes
substantially less efficient in smaller galaxies.  In halos with total
mass smaller than 10$^9$ M$_{\odot}$ (also equivalent to a virial
temperature T$_{vir}$ $<$ 10$^4$K) SN feedback and the cosmic UV
background strongly suppress SF. Furthermore, in halos this small,
stars only form when H$_2$ cooling is introduced, as gas can cool
below T$_{vir}$. As a result, in the smallest halos only a very small
fraction of baryons is then turned into stars. The more massive
galaxies in our sample turn only $\sim$ 10$\%$ of their primordial
baryon content into stars, after having expelled about 30\% of their
gas outside R$_{vir}$. Typical dwarfs in our sample turn a few per
cent of their primordial gas fraction into stars, and the smallest
galaxies $\sim$ 0.01\%. In OH11 (see their Fig.5) we verified that
galaxies with V$_{peak}<$ 60 \kms ~form the correct amount of stars
when compared with a local sample with resolved photometric and
kinematic data. In a future paper we will show how our sample closely
matches the stellar mass/halo mass relation inferred using halo
occupation methods \citep[Munshi et al. in prep.]{moster10}. As a
reference and a resolution test of the simulations, most of the above
runs have been repeated including only the collisionless CDM
component.

The DM and baryonic mass distribution of the simulated galaxies will
be compared with those measured from extensive HI data from a sample
of nearby dwarf galaxies from THINGS (Walter et al. 2008) and LITTLE
THINGS (Hunter et al. in prep) surveys which focused on field
galaxies.  The high-resolution HI data ($\sim$6$\arcsec$ angular;
$\leq$ 5.2 \kms\ velocity resolution) combined with {\it Spitzer} IRAC
3.6$\mu$m and ancillary optical images significantly reduce various
observational systematic effects inherent in lower-resolution data,
such as beam smearing, dynamical center offset and non-circular
motions, and thus enable us to derive more reliable mass models of the
galaxies.  For a comparison with our simulations, we select a sample
of 22 dwarf galaxies (7 from THINGS and 15 from LITTLE THINGS) that
show a clear rotation pattern in their velocity fields. These
high-quality multi-wavelength data allow us to measure the enclosed
amount of mass and the inner slope of the DM density profile at 500 pc
of the galaxies with good accuracy.

\section{The Evolution of DM cores as a function of halo mass and redshift}
\label{cores}

With the goal of measuring when and how much gas outflows affect the
underlying DM profiles in $\Lambda$CDM galaxies, in this section we
focus on how the central DM density profiles differs from the simple
predictions of DM-only runs once cooling, SF processes and gas
outflows are introduced. To do this we measure the slope $\alpha$ of the DM
density profile at 500pc for all the well resolved galaxies formed in
our hydrodynamical simulations, and then compare them with
observational data as well as predictions from DM--only simulations.

 The $\alpha$ value of the central DM density profile is obtained by
spherically averaging the density and fitting the density profile with
$\rho_{DM} \propto r^{\alpha}$ between 300pc and 700pc.  $\alpha$ is then
formally defined at 500pc. In this section we exclusively study field
galaxies to avoid the effects that satellite -- main halo interactions
might have on the density profiles \citep{mayer01,stoehr02,stelios04}.

\begin{figure}
\includegraphics[angle=0,width=88mm]{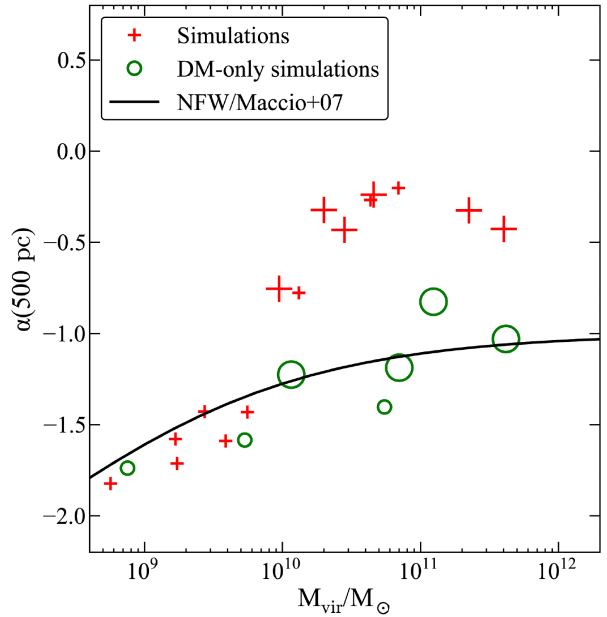}
\caption{ {\it The slope of the dark matter density profile $\alpha$
    measured at 500 pc  vs virial mass} and at z=0 for the same galaxies shown in Figure 1. 
   Crosses mark haloes from the DM$+$gas
  simulations.  Open circles are from the haloes that have been
  re--run in DM--only simulations. Size of symbols is the same as in
  Figure 1.  The solid line is the average slope predicted in Macci\'o
  \etal (2007) for haloes in the same $\Lambda$CDM cosmology.  
}
\label{fig2}
\end{figure}

\begin{figure}
\includegraphics[width=88mm]{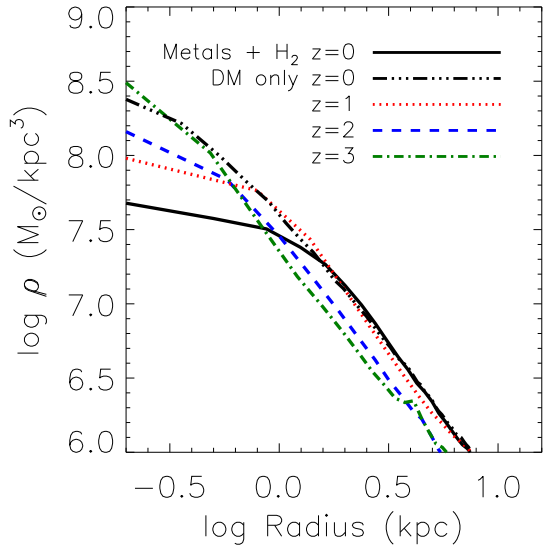}
\caption{ {\it The dark matter density
    profile of a dwarf galaxy in our sample, at $z=4,3,1,0$.} The
  prolonged process of cusp flattening due to many separate outflows
  results in a shallow inner profile at $z=0$. For comparison, the
  density profile of the same galaxy, but simulated with DM only, is
  shown in the black dash-dot line. In the DM only simulation the DM
  maintains its cuspy density profile at all redshifts.
}
\label{fig2b}
\end{figure}

Fig. 1 and Fig. 2 show the value of $\alpha$ as a function of galaxy
stellar mass and virial mass. The zoomed-in runs approach allows us
cover a large range of galaxy stellar masses, almost 6 orders of
magnitude.  Both figures clearly show a trend with increasing galaxy
stellar (or total) mass showing a central DM profile significantly
flatter than the one predicted by CDM simulations that {\it only}
included a DM component \citep[solid line, showing results
from][]{maccio07}. In Fig.1 the DM-only predictions are mapped onto
the x axis by assuming the same stellar mass - halo mass relation as
in our runs.  The DM profiles become progressively flatter up to the
most massive systems probed by our simulations, having peak velocities
of about 100 \kms. This result supports the model where SN originated
outflows are able to transfer significant amounts of energy to the DM
at the center of each galaxy, lowering the DM central density
\citep{pontzen11}. At z$=$0 our simulations predict that the central
slope of the DM density profile (again, measured between 0.3 and 0.7
kpc) as a function of stellar mass is well fitted over a mass range $4
< ~\log M_{star} < 9.4$ by:
\begin{equation}
\
\label{eq:1}
\rho_{DM} \propto r^{\alpha} ~~with~~\alpha \simeq -0.5 + 0.35 \log_{10} \left( M_{\star}/ 10^8\,M_{\odot} \right)
\
\end{equation}

\begin{figure*}
\includegraphics[width=145mm]{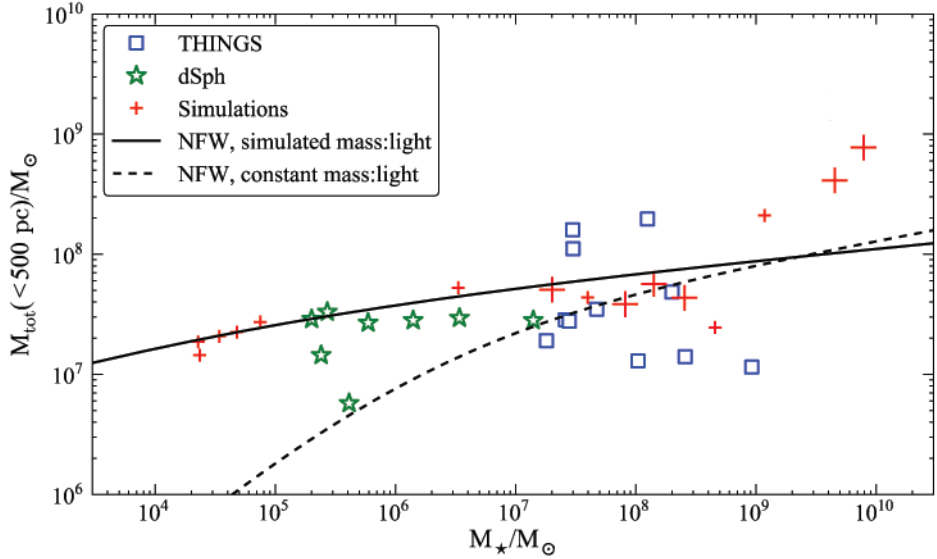}
\caption{{\it The total mass (baryons and DM) within the central 500pc
    as a function of stellar mass:} Large and small crosses:
  simulations. Open squares: galaxies from THINGS (Oh et al. in
  prep). Stars: dSph from Walker (priv. comm.).  Theoretical
  predictions reproduce the observed flat trend from 10$^5$ to 10$^9
  M\odot$.  This is largely due to the large drop in SF efficiency at
  small halo masses, that stretches the range of galaxy luminosities
  over a relatively smaller halo mass range.  The solid and dashed
  lines assume different stellar mass - total halo mass relations. A
  close fit to the simulations as M$_{\star}$ $\propto$ ${M_{Vir}}^2$
  (solid) and one showing M$_{\star}$ $\propto$ ${M_{Vir}}$
  (dashed). Only when the star formation efficiency is a steep
  function of halo mass it is possible to reproduce the observed
  trend, as discussed in \S4. More massive galaxies above the solid line have a small bulge component.}
\label{fig4}
\end{figure*}

The $\alpha$ trend in the DM-only halos in Fig. 1 and 2, with less
negative (flatter) values for increasing stellar and halo masses is
partly driven by the constant radius used to compare with observations
over a large range in galaxy masses, as in DM Einasto-like density
profiles $\alpha$ gets steeper in their outer parts (i.e for larger
R/R$_{vir}$ values), with $\alpha$ rolling from $\sim$ -1 close to the
centre to -3 at R$_{vir}$. For larger halos 500pc represents a smaller
fraction of the virial radius, hence the DM profile is flatter (less
negative $\alpha$ values) than for smaller halos, where at 500 pc we
are instead starting to measure the outer part of the DM profile.
This effect is present also in the runs with SF induced outflows, but
with a significant shift to flatter profiles and less negative
$\alpha$ values over a large range in stellar masses.  As a result
central DM profiles in runs with outflows become rapidly flatter at
halo masses larger than 10$^9$ M$_{\odot}$ (circles in Fig.2, see also
Fig.3). The DM-only runs in our sample follow closely the results from
a set of high resolution halos presented by \cite{maccio07}.  Scatter
at a similar mass is small and there is no trend with resolution
(small vs large crosses). However, in the lowest mass dwarfs in our
sample, when star formation converts less than $\simeq$ one \Msun of
baryons per 5$\times$ 10$^4$ \Msun of DM (corresponding to M$_{vir}$
$<$ 5 $\times$ 10$^9$, or also V$_{peak}$ $<$ 20 \kms) outflows are
too weak to remove sufficient amount gas and cause DM heating through
energy transfer. These halos maintain their original cuspy DM
profile. Note that halos formed in simulations that include outflows
accrete slightly less mass over a Hubble time compared to DM-only runs
(Fig.2).

In \cite{pontzen11} we presented an analytical model of the cusp
flattening process. In the galaxies in our sample this process
commonly starts at high redshift (z $\sim$ 4) and continues down to z
$\sim$ 1, when the SFR declines. Fig. 3 shows the DM (spherically
averaged) density profile as a function of radius and redshift for a
dwarf with final total mass $\sim$ 10$^{10}$ M$_{\odot}$, showing a
rapid flattening at high z.  Around z=3, as the SFR peaks, the galaxy
is undergoing several starbursts associated with a rapid accretion and
mergers. The DM density profile flattens and reaches a relatively
stable profile by z=1, when the SFR declines.  The last small
starbursts have a relatively small effect.  By z$=$0 the DM
profile shows a flattened profile within the central 1 kpc and its
central density is almost an order of magnitude lower than in the DM
only case.

An interesting result of this work is that massive blowouts, that
could potentially disrupt the formation of gaseous and stellar disks
observed in field dwarfs are {\it not} required for the formation of
cores. Each outflow typically removes gas only from a small area
(typically $<$ 1kpc) around the galaxy center and disks remain fairly
thin \citep{sanchez10}.  This model is similar to that proposed in
\cite{read05} where the need for repeated outflows was stressed, and
\cite{mashchenko08}, where ongoing SF was associated with
cusp-flattening.  In fact, some studies were not able to create cores
using single, massive blowouts \citep{gnedin02,bk11b}. Our results
suggest that those models fail as the amount of baryons removed from
the central kpc is small over a single outflow event, irrespective of
the fact that the whole galaxy gas reservoir can be blown
away. Consequently, the energy transfer to the central DM over a
single outflow event is small compared to the binding energy of the DM
cusp and multiple events become necessary.

In the scenario proposed in this work, only very small galaxies
(V$_{peak}$ $<$ 20), where SF is extremely inefficient, retain the
cuspy CDM profile that was originally predicted by DM--only
simulations. In very small objects star formation rates are too low to
generate sufficient energy to modify the CDM cusp.  This sets an
interesting scale \citep{willman05,tollerud08} where primordial cuspy
profiles would still be observable. These observations would be able
to discriminate between CDM and alternative DM models, where cores are
generated by primordial effects and would then be present even in the
lowest stellar mass objects.

The comparison with the THINGS and LITTLE THINGS dataset shown in
Fig.1 (more details in Oh et al, in prep) shows an extremely good
agreement between the simulations and the real galaxies. The
observational data set includes 22 galaxies.  The simulations
reproduce both the absolute values of $\alpha$ and the trend with
stellar mass.

Using the following formula \citep{deblok01},
  we first converted the dark matter rotation curves derived subtracting the
  baryons from the total kinematics of the 22 dwarf galaxies to the
  dark matter density profiles (see \cite{ohsim11,oh11} for more
  details),
\begin{equation}
\label{eq:2}
\rho(R) = \frac{1}{4\pi G}\Biggl[2\frac{V}{R}\frac{\partial V}{\partial R} + \Biggl(\frac{V}{R}\Biggr)^{2}\Biggr],
\end{equation}
where $V$ is the rotation velocity observed at radius $R$, and $G$ is
the gravitational constant. We then measured the logarithmic inner
slopes $\alpha$ of the derived dark matter density profiles assuming a
power law ($\rho \sim r^{\alpha}$).  After determining a break-radius
($<$ 1 kpc) where the slope changes most rapidly, we measured the inner
slope by performing a least squares fit to the data points of a given
profile within the break-radius.  A similar analysis had been
performed by OH11 on two simulated dwarfs that were compared to
galaxies from the THINGS survey. Their analysis compared estimates of
$\alpha$ obtained from the observed mass distributions and from
artificial observations (HI datacubes paired with artificial
photometric images) of the simulations. The two methods showed good
agreement.  Relevant to the results presented here, OH11 also showed
that observations can correctly recover the DM profile of galaxies
with no significant biases. Those tests support the analysis and the
results presented here.  In a future work (Oh et al in prep) the
observational dataset will be presented in more detail and it will be
compared with results obtained by artificial observations of the new
simulations.

We have verified that our results are robust versus unwanted numerical
effects. In particular we find that the graininess of the DM potential
(due to a finite number of particles) does not substantially affect
its response to gas outflows. To this aim we increased the number of
DM particles by a factor of eight for all the halos in in Field 3,
finding that the measured DM slopes remain substantially unchanged.
In G10 we verified that gas resolution effects dominate over pure DM
resolution effects as poorer resolution creates {\it more cuspy}
cores, likely as outflows become poorly resolved and artificial
viscosity brings more gas to a galaxy centre. As most halos were
re--run including only the DM component we also verified that the
central DM profiles in the absence of baryons were as cuspy as those
in the literature \citep[e.g.]{reed05a,maccio07}. Fig.2 shows good
agreement between the slopes measured from our simulations and the
predictions inferred from \cite{maccio07} over the whole sample range
in mass and resolution.  This test, combined with the existing data
points in the halo range 10$^{10}$-10$^{11}$ that span a range in mass
and force resolution (Table 1), show that core formation and sizes are
stable quantities over the resolution range explored in this work.

 The analytical model of core formation presented in
  \cite{pontzen11} shows that the creation of DM cores should be a
  generic property of fast, repeated gas (out)flows.  Core creation
  should then be a common outcome of any feedback scheme that can
  create such flows \citep{springel05,keres09,dave10,choi11,hopkins11} as
  long as sufficient spatial resolution and high SF surface densities
  are achieved.  However, as some traditional SPH implementations
  provide a poor description of Rayleigh--Taylor and Kelvin-Helmholtz
  gas instabilities \citep{agertz07} it will be important to evaluate
  how potential improvements recently introduced by different
  gasdynamical treatments \citep{read11,bauer11,keres11} affect the
  outcome of baryon energy injection from different astrophysical
  processes. Encouragingly, results showing the formation of DM cores
  due to repeated energy injection (from super massive
  black holes) into the gas component were also recently obtained
  using an AMR code \citep{martizzi11}. 

The extremely good agreement between observations and simulations
  strongly supports that baryon-DM interactions, and specifically
  repeated baryonic outflows, are able to lower the DM density at the
  center of galaxies, creating DM `cores'.  This result resolves one of the
  outstanding problems faced by the CDM model of galaxy formation,
  namely the strong discrepancy between the original model predictions
  and the observed DM distribution in galaxies.

\section{The central  mass concentration   of   galaxies:  simulations vs observations} 
\label{origin}

In this section we compare the total mass contained in the central
regions of our simulated galaxy sample with the existing constraints
coming from 12 galaxies in the LITTLE THINGS and THINGS sample and
from a sample of local dwarf spheroidals \citep[dSph]{walker09}. This
comparison is necessary to test if the distribution of stars, gas and
DM of the simulated galaxies is realistic over a range of galaxy
masses, hence making our predictions on the shape of the DM density
profiles of galaxy halos robust. With this goal we will focus on a
recent observed relation that has placed strong constraints on the CDM
model, namely the central mass -- luminosity relation \citep[S08
hereafter]{strigari08}. As our simulations have not been tuned to be a
good match to these observational data they will provide a good test
of the efficiency of SF as a function of stellar mass and of the
relative effect of outflows as a function of halo mass.

\begin{figure}
\includegraphics[trim=20mm 5mm -15mm -30mm, clip, width=75mm]{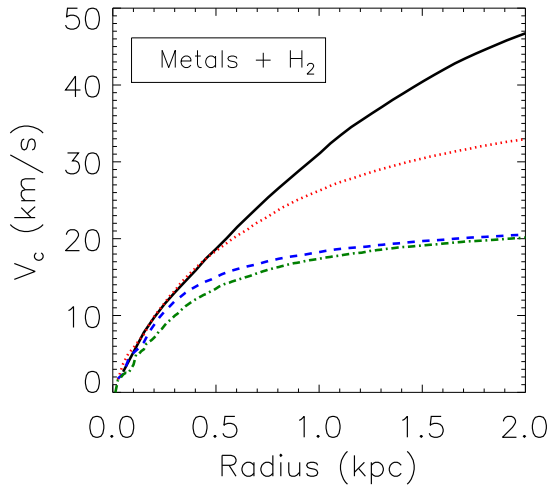}
\caption{ The $z=0$ rotation curve of a sub-sample of our simulated
  galaxies measured  at z$=$0 from the simulations that included SF and  gas processes. Each line correspond to a different galaxy. The
  amplitude of the rotation curve measured at $\sim$ 500pc is almost
  constant over a large range of masses as more massive galaxies tend
  to have flatter DM cores.  }
\label{fig5} 
\end{figure}

Fig.~\ref{fig4} shows the total mass within 500 pc for the simulated
field galaxies in our sample, as a function of their total stellar
mass. The figure also  shows similar estimates from observed galaxies. For the observational sample, stellar masses are obtained assuming a M/L $\sim$ 2 for the
Walker et al sample (Walker, priv.comm) and  from the optical and
  {\it Spitzer} IRAC 3.6$\mu$m photometry for the THINGS and LITTLE
  THINGS samples (Oh et al. in prep.). In particular, for the THINGS
  and LITTLE THINGS samples, we obtain the 3.6$\mu$m mass-to-light
  ratio values from optical colors based on \cite{bruzual03} stellar
  population synthesis models \citep{oh08,bell01}.  In
  addition, assuming a spherical potential, we calculate the total
  masses (DM + baryons) within 500 pc of the THINGS and LITTLE THINGS
  samples from their total rotation curves.

  Measurements from simulations are in extremely good agreement with
  the observed estimates, which with the addition of the THINGS and
  LITTLE THINGS datapoints extend the results in S08 by almost two
  orders of magnitude to larger galaxy masses. It is also remarkable
  that the two observational samples agree with each other very well,
  being one comprised of Milky Way satellites and the other of field,
  rotationally supported galaxies.  Our simulations predict that the
  flat central mass -- stellar mass relation will start an upward
  trend for galaxies with stellar masses $>$ 10$^9$ M$_{\odot}$. At
  that mass scale our simulated galaxies start having a small bulge
  component and the total mass within 500 pc starts growing. The flat
  central mass - stellar mass relation shown by our simulated CDM
  galaxies, differs from the naive expectations from DM--only runs,
  which show a clear correlation between halo peak velocity (or halo
  mass) and the mass within a fixed radius: in those models more
  massive halos contain more mass \citep{li09} within a fixed radius.

  Our analysis, focusing on simulated field halos, also removes the
  complications introduced by the dynamical interactions with the host
  galaxy potential as for the satellites of the Milky Way
  \citep{stoehr02}. Fig.5 shows the circular velocity V$_c$ (defined
  as $\sqrt{ (M/r)}$) for the galaxies in our sample, where M is the
  total mass within $r$. This plot supports the finding that the total
  amount of mass within the central kpc is a weak function of the
  galaxy peak velocity.  These findings suggests that in our
  simulations a flat central mass -- stellar mass relation originates
  from having a large range in luminosities over a relatively small
  range in halo masses and is possibly further helped by the
  flattening of the central DM mass profile in the more massive
  galaxies.  The continuous and dashed lines in Fig.4 show the
  resulting central mass - stellar mass relation if a different
  mapping between halo and stellar mass is adopted (normalized at
  10$^9$ M$_{\odot}$).  The continuous line is for M$_{\star}$
  $\propto$ ${M_{Vir}}^2$ as in our simulations. The dashed line
  assumes a linear relation M$_{\star}$ $\propto$ ${M_{Vir}}$ that
  would be created by having less efficient feedback at smaller
  masses, resulting in more stars. In this second case there is a
  clear trend between stellar mass and total central mass, as a much
  larger range in halo masses is mapped over the same range in
  stellar masses. As a consequence, if simulations had a weaker
  feedback they would not match the central mass - stellar mass
  relation.

  Note that we only plot the subsample of Strigari's sample that could
  be safely extrapolated out to 500pc. As shown in S08, the full
  sample continues to follow a flat central mass - stellar mass
  relation down to $\sim$~10$^3$ M$_{\odot}$ (in stars) {\it when
    measured at 300pc}.  We decided against extrapolating to 500pc the
  stellar and total masses of the smallest galaxies in the observed
  sample. This extrapolation becomes less reliable going to very
  faint/small galaxies with very small sizes (Walker, private
  comm.). On the other hand our simulations would become less reliable
  at smaller radii. We plan to study the mass distribution in
  ultra--faint dwarfs and in satellites of MW--like systems in future,
  higher resolution simulations that will allow robust predictions at
  smaller radii.

  The scatter in accretion histories and SF truncation times of the MW
  satellites due to ram pressure \citep{mayer07b} has often been cited
  as a possible origin of the central mass - stellar mass relation
  \citep{maccio09,li09,parry11}, as it would cause a large scatter in
  the amount of formed stars at a given halo mass.  This mechanism
  cannot be responsible for the flat trend in Fig.4, where the THINGS
  galaxies and all the simulated galaxies are field galaxies with a
  prolonged, untruncated SFH.  We plan a more detailed comparison
  using a larger ensemble of simulated satellites of MW analogues
  (Brooks et al, in prep.). These results are very encouraging, as the
  ability of creating realistic galaxies (see also G10) gives further
  support to gas outflows as the origin of the flattening of CDM
  density profiles.

\section{Conclusions}

In this work we used fully cosmological hydrodynamical simulations to
show that once baryonic processes are correctly taken into account,
shallow central DM profiles are a common property of field galaxies
formed within the $\Lambda$CDM model. Our predictions are in excellent
agreement with observational estimates of the DM distribution in
galaxies from the THINGS and LITTLE THINGS data samples and extend
results from \cite{G10,ohsim11}. The introduction of gas outflows in
high resolution simulations resolves the long standing tension between
the observed 'cored' DM distribution at the center of small galaxies
and the dense cuspy DM distribution predicted in (C)DM--only
simulations.  SN feedback is shown to be a vital ingredient in galaxy
formation models, where the removal of low angular momentum gas
creates {\it at the same time} galaxies with DM cores and bulgeless
dwarfs \citep[G10]{fall83,bullock01a,vdbosch01a,mashchenko08,brook11,pontzen11}.

This work highlights the fundamental role that baryon and DM
interactions play in shaping galaxy properties as fundamental as their
central DM and baryon distribution.  Predictions on the detailed
properties of galaxies based on DM--only simulations or methods where
baryon dynamics are not fully coupled to the DM need to be viewed with
some caution.

The simulated galaxies described in this work covered more than 5
orders of magnitude in stellar mass and circular velocities from 10
\kms to 100 \kms. These simulations were carried to z$=$0, included
metal cooling and H$_2$ related processes with the spatial and mass
resolution to identify individual star forming regions.  In these
simulations bursty SF limited to dense, H$_2$ rich regions creates
repeated, fast outflows which break the adiabatic approximation.  Over
several Gyrs, these fast and repeated outflows progressively lower the
central DM density of galaxy halos and turn CDM central 'cuspy'
profiles into much shallower 'cores'.

 With the
combination of SN feedback and cosmic UV background adopted in this
work, the DM distribution remains cuspy (or at least with cores
smaller than our current spatial resolution) only in dwarfs with
M$_{star}$ less than 10$^5$ M$\odot$, where less than 0.03\%
M$_{\odot}$ of the original baryon fraction was turned into stars.
These findings strongly support the analytical model presented in
\cite{pontzen11} and other numerical works on the formation of cores
\citep{G10,martizzi11,maccio11}, while making detailed observable
predictions.

As an important consistency check of our model of baryon--DM
interactions, we compared our simulations to the observed central mass
- luminosity relation for dwarf, extended to a sample containing also
filed galaxies from the THINGS and LITTLE THINGS surveys (Oh et al, in
prep, S08).  Our simulations reproduce the almost constant total mass
within the central 500pc of dwarf galaxies over a wide range of
masses, up to a few times 10$^8$ M$_{\odot}$. In our framework this
result is caused by rapidly decreasing SF efficiency at decreasing
halo masses and SN feedback simultaneously lowering the central DM
density in more massive galaxies.  These two effects cause galaxies
over a large range in luminosities to inhabit halos with a relatively
small mass range within the central kpc.

 The correct modeling of baryon--DM interactions in galaxy
  formation simulations is still in its early stages and the differences
  between various feedback schemes are possibly even larger than the
  discrepancies still existing between different gasdynamical codes
  \citep{scannapieco11}. It is remarkable that feedback
  processes similar to those commonly observed in galaxies can {\it
    simultaneously} improve on at least three fundamental problems in
  galaxy formation: i) the substructure overabundance problem by
  rapidly decreasing SF efficiency at smaller halo masses, ii) the
  formation of bulgeless galaxies in hierarchical models by the
  selective removal of low angular momentum gas and iii) the existence
  of DM cores in CDM cosmologies by allowing energy transfer from
  baryons to the DM matter.  While it is possible that other physical
  processes are involved in each of the above problems, numerical and
  analytical models point to the ubiquitous role of energy feedback.
Further comparisons with observational data, especially the
constraints coming from the Milky Way satellites, will involve
understanding in detail how the properties of the mass distribution of
faint dwarf galaxies can be inferred from their stellar kinematics
\citep{evans11,koposov11,bk11b,adams12} and how interactions with the
main galaxy affect the DM, SFHs and final stellar distribution of
galaxy satellites. We expect these comparisons to add further
constraints on the effects of SN feedback as a function of halo mass
and to guide predictions for DM direct detection experiments
\citep{dalal02}.  In the near future we will also extend our analysis
to higher mass systems as $\sim$ L$\star$ galaxies, where the presence
of DM cores is still debated \citep{swaters11,mamon11}. Measurements
of the mass distribution in very faint galaxies will be able to
strongly distinguish between baryon-DM interaction models from those
invoking alternative DM models to explain the observed central
distribution of galaxies.

\medskip
\section*{Acknowledgments}
FG and TQ were funded by NSF grant AST-0908499.  FG acknowledges
support from NSF grant AST-0607819 and NASA ATP NNX08AG84G.  AZ
acknowledges support from NSF grant AST-0908446, ISF grant 6/08 and
GIF grant G-1052-104.7/2009. AB gratefully acknowledges support from 
The Grainger Foundation. Some of SHO research was carried at "The Centre
for All-sky Astrophysics is an Australian Research Council Centre of
Excellence, funded by grant CE11E0090."  Simulations were run at TACC
and NAS. We thank Matthew Walker for sharing his data and Oleg and
Nick Gnedin, Avishai Dekel, Piero Madau, Mike Boylan-Kolchin and Jorge
Pe\~narrubia for useful discussions.

\bibliography{bibref}

\bibliographystyle{mn2e}

\bsp

\label{lastpage}

\end{document}